\documentclass[prl,twocolumn,amsmath,showpacs,superscriptaddress,floatfix]{revtex4} 
\usepackage{graphicx} 
\begin{document} 
\title{Sheared force-networks: anisotropies, yielding and geometry} 
 
\author{Jacco H. Snoeijer} 
\affiliation{Physique et M\'ecanique des Milieux 
H\'et\'erog\`enes, UMR 7636 CNRS-ESPCI, 10 rue Vauquelin, 
75231 Paris Cedex 05, France} 
 
\author{Wouter G. Ellenbroek} \affiliation{Instituut--Lorentz, 
Universiteit Leiden, Postbus 9506, 2300 RA Leiden, The Netherlands} 
 
\author{Thijs J.H. Vlugt} 
\affiliation{Department of Condensed Matter and Interfaces, 
Debye Institute, Utrecht University, P.O.Box 80.000, 
3508 TA Utrecht, The Netherlands} 
 
\author{Martin van Hecke} 
\affiliation{Kamerlingh Onnes Lab, Leiden University, PO box 9504, 
2300 RA Leiden, The Netherlands.}

\date{\today} 
 
\begin{abstract} 
A scenario for yielding of granular matter is presented by 
considering the ensemble of force networks for a given contact 
network and applied shear stress $\tau$. As $\tau$ is increased, the 
probability distribution of contact forces becomes highly 
anisotropic, the difference between average contact forces along 
minor and major axis grows, and the allowed networks span a shrinking 
subspace of all force-networks. Eventually, contacts 
start to break, and at the yielding shear stress, the packing 
becomes effectively isostatic. The size of the allowed subspace 
exhibits simple scaling properties, which lead to a prediction of 
the yield stress for packings of arbitrary contact number. 
\end{abstract} 
 
\pacs{45.70.Cc, 05.40.-a, 46.65.+g}
 
\maketitle 
 
\begin{figure}[tbp] 
\includegraphics[width=7cm]{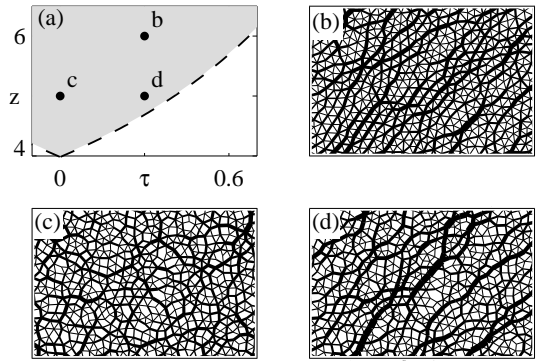} 
\caption[]{(a) Overview of the $z$-$\tau$ parameter space, where $z$ 
and $\tau$ denote coordination number and shear stress respectively. 
For a given $z$, force networks cease to exist beyond a maximum 
$\tau_m$ (dashed line). The dots indicate the values of $\tau$ and 
$z$ in panels (b-d). (b-d) Parts of the force networks, where line 
thicknesses represent the strength of the contact forces. 
These networks become highly anisotropic under shear 
stress.} \label{fig1} 
\end{figure}

Granular media, foams and emulsions are amorphous materials which 
can jam and then sustain a certain amount of shear stress 
before yielding 
\cite{jamming,ohern,radjai98,radjaisoil,herrmanisotropy}. 
If one slowly increases the applied shear stress and follows the 
evolution of contact forces and grain locations for such 
systems, one encounters a rather complex set of phenomena. Firstly, 
before the system yields as a whole, there are non-adiabatic 
precursor events such as local rearrangements and microslip 
\cite{precursor,roux,silbertprivate}. Secondly, the interparticle 
contact forces in these systems are organized into highly 
heterogeneous and fragile force networks 
\cite{gm,brujic,network,shearclement,bob} -- see Fig.~\ref{fig1}. 
Unraveling these shear-induced phenomena and their impact 
on macroscopic unjamming has remained a great challenge. 
 
While the contact forces evolve to satisfy the applied stresses, the 
selection of a {\em specific} force network for a single numerical 
experiment hinges on microscopic details and packing history. On the 
other hand, features like contact force probabilities have turned 
out to be relatively insensitive to these subtle details 
\cite{network,brujic,grestjam,bob,corwin}, which suggests a 
purely statistical approach. In this Letter we characterize 
granular packs under shear stress, by studying 
{\em ensembles} of force networks for fixed contact networks 
\cite{jp,prlensemble,PRET,Unger,herrmannensemble,socolar}. 
This approach is based on the fact that in jammed systems there are 
more contact forces than force balance equations -- 
the ensemble simply consists of all those force networks 
for which the contact forces are repulsive, balance on every grain 
and satisfy global stress constraints, while keeping the geometry fixed. 
This provides a novel access to the statisics of force networks under 
shear stress, in which the roles of fabric and force anisotropy 
are separated explicitly. 
 
We consider ensembles of sheared force networks for frictionless 
disks in two dimensions, for contact numbers $z$ ranging from the 
lower limit $z\!\approx\! z_c\!=\!4$ (isostatic \cite{isostatic}) to 
$z\!=\!6$ (strongly hyperstatic). We recover a number of 
experimental features, in particular a transition to yielding: force 
networks cease to exist beyond a critical yield stress $\tau_m$. The 
boundary of the grey zone in Fig.~1a indicates that this maximum 
yield stress strongly depends on the coordination number. 
 
In this paper we characterize the physics of sheared force networks 
at three levels of detail. {\em{(i)}} 
$P_{\phi}(f)$, the contact-angle ($\phi$) resolved probability 
density for the contact forces $f$, 
is a natural extension of the overall force distribution $P(f)$ 
\cite{network,brujic,grestjam,bob,corwin}, 
and acts as a sensitive probe for anisotropy due to shear. 
{\em{(ii)}} The angle resolved {\em 
average} force $\bar{f}(\phi)$ is possibly the simplest characterization 
of shear-induced anisotropy. We find that it evolves sinusoidally as 
$\bar{f}(\phi) =1 + 2 \tau \sin(2 \phi)$ for small $\tau$, but that 
higher harmonics develop for larger shear stresses. Bounds on $\tau_m$ 
are obtained by requiring that $\bar{f}(\phi)$ should not become 
negative.  {\em{(iii)}} The yielding curve of Fig.~1a is due to 
a vanishing volume of the high-dimensional space of allowed 
networks ${\bf F_\tau}$. Characterizing the size of ${\bf F_\tau}$ by the average 
distance $L$ between random pairs of force networks, we find that 
$L$ and $\tau$ are approximately related as $(L/L_m)^2 + (\tau/\tau_m)^2 =1$. 
Both $L_m$ and $\tau_m$ exhibit simple scaling laws with contact number $z$, 
$z-z_c$ and particle number $N$, resulting in a general prediction 
for the maximal shear stress $\tau_m$ that a given granular 
packing can sustain before yielding.

\begin{figure}[tbp] 
\includegraphics[width=7cm]{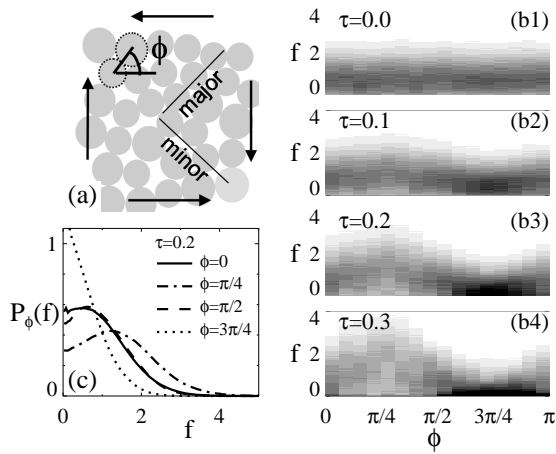} 
\caption[]{(a) Illustration of our geometry, showing the direction 
of shear stress (arrows), contact angle $\phi$ and major and minor 
axes. (b) Angle resolved force distributions for $z\!=\!5$. 
$P_{\phi}(f)$ becomes increasingly modulated with contact angle 
$\phi$ for increasing shear stress $\tau$. (c) $P_{\phi}(f)$ varies 
qualitatively with $\phi$ (here $\tau\!=\!0.2$).} 
\label{fig2} 
\end{figure} 
 
{\em Force ensemble -- } The numerical results have been obtained by 
a recently developed ensemble technique 
\cite{prlensemble,PRET,herrmannensemble,Unger,socolar}. The input of the ensemble 
consists of a fixed contact geometry of a 2D packing of $N=1024$ 
frictionless disks of radii $R_i$ with centers ${\bf r}_i$ and 
coordination number $z>z_c$ in a volume $V$, which we generated from 
molecular dynamics simulations of a 50:50 binary mixture of 
particles with size ratio 1.4 that have a purely repulsive 
Lennard-Jones interaction. Different densities were used to obtain 
different $z$. The system is {\em not} sheared and the resulting 
contact networks are isotropic \cite{ohern,prlensemble}. These 
contact networks are then kept fixed, and the positive interparticle 
forces between particles $i$ and $j$, $f_{ij}$, are seen as degrees 
of freedom which satisfy mechanical equilibrium, restricted by the 
macroscopic stress $\sigma_{\alpha\beta}$: 
\begin{equation}\label{mecheq} 
\sum_j f_{ij}\,\frac{{\bf r}_i-{\bf r}_j}{|{\bf r}_i-{\bf r}_j|} 
={\bf 0}~,~~\sigma_{\alpha\beta} = \frac{1}{V}\,\sum_{\{ ij\}} 
\left( {\bf f}_{ij}\right)_\alpha \left( {\bf r}_i-{\bf 
r}_j\right)_\beta~. 
\end{equation} 
In this picture there are $(zN/2)$ degrees of freedom (contact 
forces) constrained by $(2N+3)$ equations, leading, for $z\!>\!4$, to 
an {\em ensemble of force networks} that form a high-dimensional 
convex subspace ${\bf F_{\tau}}$ 
\cite{prlensemble,PRET,Unger}. In this paper we choose the 
coordinates and pressure such that 
$\sigma_{xx}\!=\!\sigma_{yy}\!=\!1/2$, and consider the 
dimensionless shear stress $\tau=\sigma_{xy}/\sigma_{xx}$ 
(equivalent to the relative deviatoric stress) (see Fig.~2a). 
 
\begin{figure}[tbp] 
\includegraphics[width=7cm]{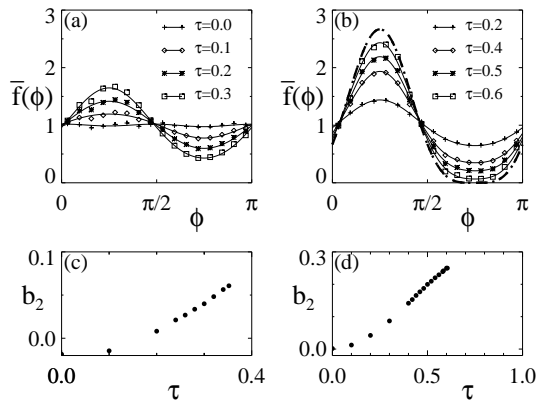} 
\caption[]{Average contact force as function of contact angle $\phi$ 
and applied shear stress $\tau$. (a) $\bar{f}(\phi)$, for $z\!=\!5$ 
develops a sinusoidal modulation when $\tau$ is increased. 
$\bar{f}(\phi)$ can be fitted well by an expression of the form 
$1+2\tau \sin(\phi) -b_2 \cos(4 \phi)$; $b_2$ slowly increases with 
$\tau$ as shown in panel c. (b) $\bar{f}(\phi)$, for $z\!=\!6$ 
develops a strong second harmonic for large $\tau$ and approaches 
the limiting curve given by Eq.~(\ref{limit}) (dashed curve). (c-d) 
Corresponding dependence of $b_2$ on $\tau$.} \label{fig3a} 
\end{figure} 
 
{\em Angle resolved force distributions -- } To characterize the 
anisotropic force networks, we introduce here the contact-angle 
resolved force distribution $P_{\phi}(f)$. This distribution is 
normalized such that  $\int df P_{\phi}(f) \!=\!1$ for all $\phi$, 
and $1/\pi \int d\phi \int df \left[f P_{\phi}(f)\right]\!=\!1$. 
Note that, in general, the average force for contacts {\em in a 
certain direction}, $\bar{f}(\phi) \equiv \int df\left[f 
P_{\phi}(f)\right]$, is {\em not} equal to one. 
 
Figure~2b illustrates that for sheared systems $P_{\phi}(f)$ modulates with 
$\phi$ -- this modulation has its extrema along major and minor axes. 
Many contact forces along the minor axis ($\phi  \sim 
3\pi/4$) evolve towards zero for increasing shear, effectively 
breaking these contacts \cite{prlensemble} and leading to a 
$\delta$-peak at $f\!=\!0$ (black area in Fig.~2b4). Fig.~2c shows 
examples of $P_{\phi}(f) $ and illustrates that for sheared systems, 
the qualitative {\em shape} of the force distribution varies with 
contact angle. In particular for large stresses, $P_{\phi}(f) $ 
becomes completely monotonic along the minor axis but remains peaked 
along the major axis. 
Earlier experimental and numerical findings that the total force 
distribution, averaged over all orientations, evolves from a peaked 
to a monotonically decreasing function with increasing $\tau$ are 
thus seen to result from averaging over strongly different 
$P_{\phi}(f)$ \cite{jamming,ohern,grestjam,bob,corwin}.

{\em Analytical bounds on $\tau$ -- } The most basic manifestation 
of stress anisotropy, however, is the modulation of the average 
force, $\bar{f}(\phi)$, as a function of the contact orientation. 
This effect is clearly visible in Figs.~1, and has only recently 
been accessed experimentally \cite{shearclement,bob}. In Fig.~3 we 
therefore show examples of $\bar{f}(\phi)$ for various stresses and 
contact numbers, as obtained by the ensemble. For the strongly 
hyperstatic case (Fig.~3b,d), it is clear that the anisotropy is 
limited by the requirement that $\bar{f}(\phi)$ should definitely 
remain positive for all $\phi$. This is due to the repulsive nature 
of the contact forces, which requires all $f_{ij}\geq0$. We show 
below that this simple criterion imposes analytical bounds on the 
maximum shear stress $\tau_m$ that become increasingly accurate for 
strongly hyperstatic packings. 
 
This bound can be computed from the probabilistic version 
of Eq.~(\ref{mecheq}) for isotropic contacts, 
\begin{eqnarray}\label{sigmaxxnofriction} 
\sigma_{\alpha\beta} &=&  \frac{{\bar r} N_c}{V\pi} \, \int_0^{\pi} 
d\phi \,  \bar{f}(\phi) \, n_\alpha \, n_\beta~, 
\end{eqnarray} 
where  ${\bar r}$ is the average particle radius \cite{footradius}, 
while $(n_x,n_y) = (\cos \phi,\sin\phi)$; in the remainder we set 
the prefactor $ {\bar r} N_c/V$ equal to unity. 
We Fourier expand $\bar{f} (\phi)$ as 
$ \sum \,a_k\sin 2k\phi + \sum \, b_k \cos 2k\phi$, 
and anticipate that only odd $k$ sine terms and even 
$k$ cosine terms are compatible with the symmetry of a simple shear 
\cite{radjai98}. 
Eq.~(\ref{sigmaxxnofriction}) then yields 
\begin{equation}\label{series} 
\bar{f}(\phi) = 1 \,+ \, 2\tau \sin 2\phi  \,-\, b_2 \cos 4\phi \,+ 
\cdots~, 
\end{equation} 
where the coefficients of the higher order terms $b_2,\dots$ are 
independent of the stress tensor (Fourier modes with $k\geq 2$ yield 
zero upon integration). The role of the higher order terms is 
limited \cite{notehot}: our numerics show a significant contribution 
to $\bar{f}(\phi)$ for large $\tau$ and $z$ for $k=2$ only 
(Fig.~3b,d).  Truncating Eq.~(\ref{series}) at second order yields 
that an upper bound is reached for $\tau\!=\! \tau_m\!=\!2/3$: 
\begin{equation}\label{limit} 
\bar{f}_{max}(\phi) = 1 + 4/3 \sin 2\phi  \,-\, 1/3\cos 4\phi 
\end{equation} 
By extending the expression 
(\ref{sigmaxxnofriction}) to include fabric anisotropy and frictional 
forces, and following a similar line of reasoning, bounds can also 
be obtained for the completely general case \cite{inprep}. 
 
Fig.~3b,d illustrate the relevance of this bound for strongly 
hyperstatic packings: for $z\!=\!6$ the maximal stress is close to 
2/3, while $\bar{f}(\phi)$ approaches the limiting form 
Eq.~(\ref{limit}), indicated by the dashed line. Closer to 
isostaticity, however, the system has much less freedom to 
accommodate to the shear stress, and $\tau_m$ is much smaller -- as 
shown in Fig.~3a,c, $\bar{f}$ then does not approach zero yet. 
In the isostatic limit there are no adjustable degrees of freedom 
left, the force space ${\bf F_{\tau}}$ shrinks to a single point, 
and $\tau_m$ tends to zero. 
In general, the question is thus how $\tau_m$ depends on $z$ (Fig.~1a).
 
{\em High-dimensional ensemble -- } The yielding curve 
$\tau_m(z)$ can be determined best from the geometric properties of 
this high-dimensional force space. We have quantified the "size" of 
${\bf F_{\tau}}$ by the Euclidian distance between random 
realizations (force networks), for given $z$ and $\tau$ 
\cite{Unger,herrmannensemble}. While the distances between random 
points in a low dimensional space are broadly distributed, this 
distribution becomes increasingly sharply peaked for higher 
dimensional objects as is the case here. The average distance $L$ 
defined via 
\begin{equation} 
L^2(z,\tau)\equiv \left< \Sigma_{ij} (f_{ij} - f'_{ij})^2 \right>~, 
\end{equation} 
thus serves as an effective measure for the size of ${\bf F_{\tau}}$. 
Here the brackets denote an average over the random pairs of 
force networks $\{f_{ij}\}$ and $\{f'_{ij}\}$ 
 
\begin{figure}[tbp] 
\includegraphics[width=7cm]{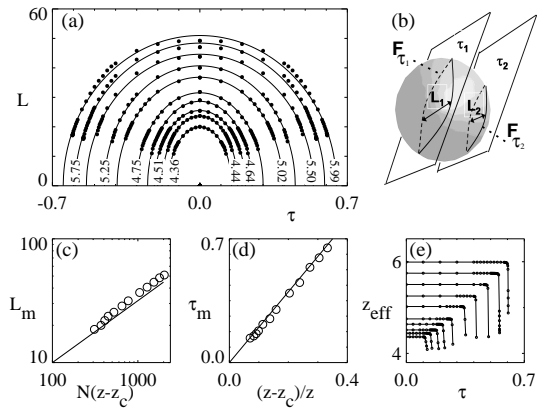} 
 \caption[]{(a) The linear size $L$ of the force space ${\bf 
F_{\tau}}$ as a function of shear stress $\tau$ for $z$ as labeled 
(dots) are well fitted by a relation of the form 
$(L/L_m)^2+(\tau/\tau_m)^2\!=\!1$ (solid curves). (b) 
 Sketch of how force space ${\bf F}$ (represented schematically by a sphere) 
 intersected by the 
 constant-$\tau$ hyperplanes yields the force spaces ${\bf 
 F}_{\tau_1}$ and ${\bf 
 F}_{\tau_2}$ with sizes $L_1$ and $L_2$. 
 (c) $L_m$ obtained by fitting Eq.~(\ref{ellipse}) (circles), 
compared to $\sqrt{N (z-z_c)}$ (line). (d) Maximal shear stress 
$\tau_m$ obtained by fitting Eq.~(\ref{ellipse}) (circles) is 
well approximated by $2 (z-z_c)/z$ (line). 
(e) $z_{\mathrm{eff}}$ drops sharply when approaching $\tau_m$. 
 } \label{fig4} 
\end{figure} 
 
In Fig.~4 we show our main findings for the main properties of the 
force space with $\tau$. First, $L(\tau,z)$ can, surprisingly, be 
fitted by a simple relation of the form 
\begin{equation}\label{ellipse} 
(L/L_m)^2+(\tau/\tau_m)^2=1~, 
\end{equation} 
which becomes particularly accurate for $z<5$ (Fig.~4a). In 
addition, as shown in Fig.~4c-d, $L_m$ and $\tau_m$ approach 
simple scaling laws: 
\begin{eqnarray}\label{scl2} 
\tau_m &\approx& 2 \,\frac{z-z_c}{z}~, \\ 
L_m &\approx& \sqrt{N(z-z_c)}~.\label{scl} 
\end{eqnarray} 
 
The scaling relations Eqs.~(\ref{ellipse}-\ref{scl}) can be 
interpreted geometrically, keeping in mind that high-dimensional 
objects can be quite counter-intuitive (Fig.~4b). Let us consider 
the ensemble ${\bf F}$ that is obtained by applying all force 
balance equations and $\sigma_{xx}\!=\!\sigma_{yy}\!=\!1/2$, but 
leaving $\sigma_{xy}$, and thus $\tau$, undetermined. ${\bf F}$ is a 
convex body of dimension $D=(z-z_c)N/2-2$ that has $zN/2$ facets; 
each facet corresponds to a certain contact force being zero. One 
can obtain ${\bf F_{\tau}}$ from intersection of ${\bf F}$ with the 
codimension-one hyperplane given by the linear constraint 
$\sigma_{xy} = \tau/2$ (Eq.~\ref{mecheq}). Due to symmetry, the most 
``central'' intersection is obtained for $\tau=0$, while for larger 
values of $\tau$, the intersection is less centered and ${\bf 
F}_{\tau}$ is smaller. 
 
In fact, Eq.~(\ref{scl}) can be recovered analytically. The scaling 
$L_m \propto \sqrt{D}$ with the dimension $D\!=\!N/2(z-z_c)$ 
of ${\bf F_{\tau}}$ is a common feature of high-dimensional convex 
spaces \cite{cubenote}. If we consider a simple approximation for 
${\bf F}_{\tau}$ by ignoring the force balance equations but only 
requiring that all $f_{ij}$ are positive and have $\langle f_{ij} 
\rangle=1$, we obtain that $L^2 \!=\! 2D \!=\! N(z-z_c)$ 
\cite{boltzmannote}. Note that the relation $\tau_m \approx 2 
(z-z_c)/z$, implies that $\tau_m$ is proportional to the ratio 
between the dimension of ${\bf F_{\tau}}$ and the number of its 
facets $zN/2$. We have not been able to come up with a convincing 
argument for this scaling, however. 
 
Finally, as $\tau$ approaches its maximum $\tau_m$, the space ${\bf 
F_{\tau}}$ shrinks and $L \rightarrow 0$, so that at 
$\tau\!=\!\tau_m$, ${\bf F_{\tau}}$ consists of a single point. The 
effective contact number $z_{\mathrm{eff}}$, which is defined by 
considering contacts broken when their force drops below a fixed 
small threshold \cite{footcutoff}, stays constant over most of the 
range of $\tau$, but sharply drops to $z_c$ as $\tau$ approaches 
$\tau_m$ (Fig.~4e). 
 
\paragraph{Outlook --} 
 
We have proposed an ensemble theory for force networks under shear, 
by exploring solutions of mechanical equilibrium for packings of 
fixed contact geometry. The ensemble approach provides a great 
conceptual simplification with respect to full numerical 
simulations, as it steps aside the intricate evolution of the 
contact network. Yet, it captures recently measured statistical 
properties, such as $\bar{f}(\phi)$ and the evolution of $P(f)$ 
\cite{shearclement,bob}. Furthermore, it provides an alternative 
description of yielding phenomena, in terms of a vanishing volume of 
the force phase space. Consistent with existing numerical 
simulations \cite{radjaisoil,roux}, we found that the maximum shear 
stress $\tau_m$, strongly depends on the coordination number of the 
packing. This dependence can be understood in terms of the geometry 
of the force space, and obeys a simple scaling law Eq.~(\ref{scl2}). The 
ensemble thus provides a new perspective for soil mechanics, in 
which relations between the macroscopic effective friction and 
micromechanical properties (density, coordination number, 
texture,... ) play a central role. 
More generally, it suggests a route along which the unjamming by shear of 
a broad range of disordered media may be understood. 
 
{\em Acknowledgements} We thank Alexander Morozov and Wim van Saarloos 
for discussion. JHS acknowledges financial support by a Marie Curie 
European Fellowship FP6 (MEIF-CT2003-502006), WGE from the physics foundation FOM, and TJHV and MvH from the science foundation NWO through VIDI grants.


\begin{thebibliography}{99}
\bibitem{footradius} Numerical simulations \cite{radjai98,prlensemble} 
have shown that $r$ ($\approx$ grain size) does not correlate to 
forces and angles, so it can be taken out of the integral.
\bibitem{notehot} ${\bar f}(\phi)$ is monotonic between the major and 
 minor principal directions, and for particles with $z$ contact 
 forces, the width of the modulation is bounded from below by roughly 
 $2\pi/z$ \cite{radjaisoil,checkrad}.
\bibitem{cubenote} For example, for a hypercube of dimension $D$ and edge $2$ 
(such that the average value of any of its coordinates is one, to 
normalize similar to the forces), $L^2 =(D/4 \int_0^2 dx \int_0^2 
dx' (x-x')^2 )=2D/3$.
\bibitem{boltzmannote} This space is exactly that of the micro-canonical 
ensemble for a finite number of forces, for which $P(f)=e^{-f}$. For 
such a space dimension of $D$, we find that $L^2 = D \int_0^\infty 
dx \int_0^\infty dx' (x-x')^2 \exp(-x) \exp(-x')=2D$.
\bibitem{footcutoff} Numerically, we consider a contact as broken when $f < 
10^{-4}\langle f \rangle$, and we have verified that results are 
independent of this cutoff.
\end{thebibliography}
\end{document}